\documentclass[12pt]{article}
\input epsf
\newcommand{\be}{\begin{equation}}
\newcommand{\ee}{\end{equation}}
\newcommand{\bea}{\begin{eqnarray}}
\newcommand{\eea}{\end{eqnarray}}
\newcommand{\ba}{\begin{array}}

\newcommand{\ea}{\end{array}}

\def\bbox{{\,
\lower0.9pt\vbox{\hrule \hbox{\vrule height 0.2 cm
\hskip 0.2 cm \vrule height 0.2 cm}\hrule}\,}}
\newcommand{\dsl}{\pa \kern-0.5em /}

\newcommand{\nn}{\nonumber \\}

\def\ds{\raise.15ex\hbox{/}\kern-.57em\partial}
\def\Ds{\,\raise.15ex\hbox{/}\mkern-13.5mu D}
\begin{document}

\baselineskip 18pt


\begin{titlepage}
\vfill
\begin{flushright}
KIAS-P01024\\
hep-th/0106165\\
\end{flushright}

\vfill

\begin{center}
\baselineskip=16pt 
{\Large\bf Anomaly of $(2,0)$ Theories} \\
\vskip 10.mm
{Piljin Yi $^\star$}
\vskip 0.5cm
{\small\it
School of Physics, Korea Institute for Advanced Study\\
207-43, Cheongryangri-Dong, Dongdaemun-Gu, Seoul 130-012, Korea}
\end{center}
\vfill
\par
\begin{center}
{\bf ABSTRACT}
\end{center}
\begin{quote}
We compute gravitational and axial anomaly for D-type
$(2,0)$ theories realized on $N$ pairs of coincident M5-branes at $R^5/Z_2$ 
orbifold fixed point. We first summarize work by Harvey, Minasian, and Moore 
on A-type $(2,0)$ theories, and then extend it to include the 
effect of orbifold fixed point. The net anomaly inflow follows when we further
take into account the consistency of $T^5/Z_2$ M-theory orbifold. We deduce
that the world-volume anomaly is given by $N{\cal J}_8 + N(2N-1)(2N-2)
p_2/24$ where ${\cal J}_8$ is the anomaly polynomial of a free tensor 
multiplet and $p_2$ is the second Pontryagin class associated with the
normal bundle. This result is in accord with Intriligator's conjecture.
\end{quote}

\vfill
\vskip 5mm
\hrule width 5.cm
\vskip 5mm
\begin{quote}
{\small
\noindent $^\star$ E-mail: piljin@kias.re.kr\\
}
\end{quote}
\end{titlepage}
\setcounter{equation}{0}

\section{Introduction}

One of more mysterious results from study of string theory, is existence
of nontrivial quantum theories in six dimensions. One class of these are
known simply as $(2,0)$ theories, which are supposed to be a non-Abelian
generalization of free noninteracting tensor theories. Furthermore, there 
are  three different types of $(2,0)$ theories, classified
by the ubiquitous ADE classification. 
One way \cite{witten6d} of realizing these theories
is to consider type IIB compactification on an ALE space that asymptotes to
$R^4/\Gamma$ where $\Gamma$ is one of finite subgroups of $SU(2)$, and
take a small coupling limit of the string theory, while collapsing the
cycles in the ALE spaces appropriately. Chiral 2-form tensors 
arise from chiral  4-form tensor of type IIB theories, while
D3-branes wrapped on the collapsing cycles provide ``charged'' degrees of
freedom that are necessary to complete the $(2,0)$ theories. The
ADE classification of the $(2,0)$ theories follows from ADE classification
of $\Gamma$.

In part 
because these theories cannot be written down in terms of usual path integral
of local fields, very little is known about them. 
On the other hand, the question of gravitational and axial anomaly is 
inherently topological, and should be accessible without detailed knowledge
of the theory. Indeed the question of gravitational and axial anomaly
of these theories has a close tie to various topological terms in M-theory,
in much the same way that the anomaly structure of heterotic string theory
was vital in understanding topological aspects of M-theory 
\cite{horava1,horava2}. For the purpose, one considers $N$ coincident M5 
branes \cite{andy6d} in the limit of divergent Planck scale.
In the flat space-time background, the tensor multiplets living on the
M5 branes, along with degrees of freedom from open membranes suspended between
adjacent M5, constitute $A_{N-1}$ $(2,0)$ theory plus a free $(2,0)$ 
tensor theory. To produce $D$-type, we replace the flat space-time with an 
orbifold $R^{1+5}\times R^5/Z_2$, and put M5's at the origin.  

In this latter context, anomaly of $(2,0)$ theories can be computed indirectly
by asking what is the anomaly inflow from bulk onto M5 brane system; the 
total anomaly should vanish since we expect M-theory with M5 brane to be
self-consistent. For the case of a single M5 brane in flat spacetime, 
Freed, Harvey, Minasian and Moore (FHMM) \cite{fhmm} observed  that a 
particular deformation of the cubic Chern-Simon term of M-theory,
\be
\int C\wedge dC\wedge dC,
\ee
appears naturally and seems necessary to cancel a single M5 brane anomaly
\cite{M5} completely. 

Subsequent extrapolation to many coincident M5's, due to Harvey, Minasian 
and Moore (HMM) \cite{hmm} allows a simple computation of the anomaly 
polynomial of A-type theory.
Eleven dimensional Lorentz group descends down
to the six-dimensional Lorentz group and the axial $SO(5)$ group, each
becoming structure groups of tangent and normal bundles of the M5's,
respectively. In this unified Language, the total anomaly polynomial
was argued to be,
\be
N{\cal J}_8+(N^3-N)\,\frac{p_2({\cal N})}{24},
\ee
where ${\cal J}_8$ is the one-loop anomaly polynomial of a single 
$(2,0)$ tensor multiplet, while $p_2$ is the second Pontryagin class 
of the normal bundle. After removing contribution from the free, ``center 
of mass'', (2,0) tensor multiplet, one finds
\be
(N-1){\cal J}_8+(N^3-N)\,\frac{p_2({\cal N})}{24},
\ee
as the anomaly of $A_{N-1}$ theory.

In this paper, we compute the anomaly polynomial of
for D-type $(2,0)$ theories by computing net anomaly inflow from bulk
onto $N$ pairs of coincident M5 branes at $R^5/Z_2$ orbifold fixed point. 
We will argue that the anomaly polynomial in this case is,
\be
N{\cal J}_8+N(2N-1)(2N-2)\,\frac{p_2({\cal N})}{24}.
\ee
The method here relies on an extension  of FHMM combined with 
the consistency of $T^5/Z_2$ compactification of M-theory.

In section 2, we review  anomaly cancellation for an
M5-brane in flat space-time, and then summarize  FHMM. We close
the section with HMM anomaly computation of A-type theories.
In section 3, we generalize to the D-type theory
by introducing an orbifold of the form $R^5/Z_2$.
We propose a simple extension of FHMM to backgrounds with orbifold
point, and point out how the anomaly inflow mediated by the antisymmetric
tensor charge would be modified. Furthermore, we point out that
there is additional
inflow associated with the orbifold singularities. We deduce this last 
contribution from the consistency of the $T^5/Z_2$ compactification,
and compute the anomaly of D-type $(2,0)$ theories. 
In section 4, we offer an independent check by estimating the one-loop anomaly
contribution from the untwisted sector. It turns out to be
consistent with the procedure we adopted,
modulo a term that could be cancelled by a six-dimensional
local counter-term. In the final section, we discuss a recent conjecture 
by Intriligator on general form of anomaly of all ADE $(2,0)$ theories,
and close with a few related comments.

\section{Anomaly Inflow onto M5 Branes}

Let us define an eight-form character $I_8$ of the space-time curvature
$R$,
\be
I_8\equiv -\frac{1}{48}\left(p_2 -\frac{p_1^2}{4}\right),
\ee
where $p_n$ is the n-th Pontryagin character. $I_8$
appears in numerous context in superstring theories and M-theory, but 
one directly relevant to us is the topological coupling that exists in
M-theory \cite{duff},
\be
\int C\wedge I_8 =\int G\wedge I_7^{(0)}.
\ee
$C$ is the 3-form tensor, $G$ is the field strength of $C$, 
and $dI_7^{(0)}=I_8$.

Consider an 
M5 brane in flat space-time. Because M5 is a magnetically charged object,
\be
dG=2\pi\delta_{\rm M5},
\ee
there is a six-dimensional gravitational anomaly, generated by variation 
of the above coupling, 
\be
-2\pi \int\delta_{\rm M5}\wedge I_6^{(1)}=-2\pi \int_{\rm M5} I_6^{(1)},
\label{inflow}
\ee
with $dI_6^{(1)}=\delta I_7^{(0)}$. On the other hand,
a single M5 composed of a single tensor multiplet of $(2,0)$ supersymmmetry.
This field content is anomalous and generate one-loop gravitational 
anomaly of amount
\be
2\pi \int_{M5} J_6^{(1)} \label{wv}
\ee
where  $dJ_6^{(1)}=\delta J_7^{(0)}$, $dJ_7^{(0)}=J_8$, and
\be
J_8= -\frac{1}{48}\left(p_2({\cal T}) -\frac{p_1^2({\cal T})}{4}\right).
\ee
where the characters are those of the tangent bundle $\cal T$ of the
six dimensional worldvolume. 
This anomaly polynomial is equal to $I_8$ provided that the latter
is also evaluated for ${\cal T}$.  The resulting anomaly is purely
gravitational in six dimensions, and the two clearly cancel each other.
We learn that $C\wedge I_8$ is essential in establishing consistency
of M5 branes in M theory.

However, not all eleven-dimensional gravitational anomaly would be cancelled
this way. Following \cite{M5}, we  split
the spacetime curvature $R$ into two parts; one associated 
with the six-dimensional tangent bundle $\cal T$ 
of M5 and the other associated with
the five-dimensional normal bundle $\cal N$ of M5.
On closer inspection, the actual anomaly
inflow from $C\wedge I_8$ term corresponds to an 
anomaly polynomial of the form
\bea
-I_8({\cal T}\oplus{\cal N})&=&\frac{1}{48}\left(p_2({\cal T}\oplus {\cal N}) 
-\frac{p_1^2({\cal T}\oplus {\cal N})}{4}\right)\nn
&=&\frac{1}{48}\left(p_2({\cal T})+p_2({\cal N}) 
-\frac{(p_1({\cal T})-p_1({\cal N}))^2}{4}\right). 
\eea
We used the fact, $p_2({\cal T}\oplus {\cal N})
=p_2({\cal T})+p_2({\cal N})+p_1({\cal T})\wedge p_1({\cal N})$.
The structure group $SO(5)$ of the normal bundle acts as the R-symmetry
group in six-dimensional theory, so the part dependent on $\cal N$
generates the axial anomaly. 

However, the anomaly associated
with a single tensor multiplet is found to be,
\be
{\cal J}_8\equiv
-\frac{1}{48}\left(p_2({\cal T})-p_2({\cal N}) 
-\frac{(p_1({\cal T})-p_1({\cal N}))^2}{4}\right),
\ee
and we  have uncanceled anomaly of the amount \cite{M5},
\be
-I_8({\cal T}\oplus{\cal N})+{\cal J}_8
=\frac{1}{24}\,p_2({\cal N}), \label{sum}
\ee
which is purely axial. Since the
axial group  $SO(5)$ also came from space-time symmetry,  its anomaly
must cancel in the full M-theory context.

An elegant solution to this puzzle has been proposed by FHMM. 
They considered a Chern-Simons term,
\be
-\frac{2\pi}{6}\int C\wedge dC\wedge dC,
\ee
argued that this coupling must be modified in a very specific manner
in the presence of magnetic sources to $C$. Introducing the Thom class $e_4$
for the normal bundle $\cal N$ of the magnetic source, they smeared out the
magnetic source by writing
\be
dG=d\rho(r)\wedge e_4/2,
\ee
where the one-form $d\rho$ is essentially the distribution of the smeared
out magnetic source \cite{fhmm}. This allows a simple modification 
of the above coupling to 
\be
-\frac{2\pi}{6}\int \tilde C\wedge d\tilde C\wedge d\tilde C,
\ee
with $\tilde C=C-\rho e_3^{(0)}/2 $, while the gauge invariant, {\it 
everywhere finite}, field strength $G$ is also modified to 
\be
G=dC-d\rho\wedge e_3^{(0)}/2=d\tilde C +\rho e_4/2.
\ee
Gauge transformation of $e_3^{(0)}$ induces that of $\tilde C$,
\be
\delta \tilde C= -d(\rho e_2^{(1)})/2,
\ee
which leads to another anomaly inflow, 
\be
-\frac{2\pi}{6}\;\delta \int \tilde C\wedge d\tilde C\wedge d\tilde C
\ee
The integrand is exact, so the integral reduces to the boundary which
is the infinitesimal sphere bundle enclosing the M5 brane. Because the
boundary involves the infinitesimal sphere, the integral actually reduces
to that over M5 brane. Results
by Bott and Cattaneo \cite{bott} show that the resulting anomaly inflow
corresponds to the anomaly polynomial,
\be
-\frac{1}{24}\,p_2({\cal N}).
\ee
Note that this precisely cancels the remaining anomaly. 
This mechanism was further
studied by Becker and Becker \cite{beckers} who reconciled it with the
cancellation mechanism suggested by Witten \cite{M5} in type 
II perspective, while
Ref.~\cite{harvey} made an attempt to study the origin of the function
$\rho$.

Subsequently, Harvey, Minasian, and Moore (HMM) observed \cite{hmm}
that this last 
anomaly inflow from Chern-Simons term is cubic in the magnetic charge, 
so the net anomaly inflow onto $N$ coincident M5 branes must be
\be
\left(-NI_8({\cal T}\oplus {\cal N})-N^3\;\frac{p_2({\cal N})}{24}\right)=
-\left(N{\cal J}_8+(N^3-N)\frac{p_2({\cal N})}{24}\right),
\ee

\section{Anomaly of D-type Theories}

To study D-type $(2,0)$ theory, it suffices to put $2N$ coincident M5-branes 
at origin of $R^5$ and orbifolding the latter by the parity $Z_2$,
\be
x^i\rightarrow -x^i\qquad i=6,7,8,9,11.
\ee
This $Z_2$ has to act on the antisymmetric tensor $C$ as
\be
C\rightarrow -C,
\ee
upon which both topological terms of the previous section remain
invariant. 

Since this background is not a string theory vacuum, the role of orbifold
fixed point is less transparent. Nevertheless, there are several 
facts known about it. First of all, the orbifold preserves the same
set of supersymmetry preserved by M5 branes, so that the world-volume
supersymmetry remains $(2,0)$ on any M5 transverse to $R^5/Z_2$.
It is also known that the twisted sector fields
can be attributed to world-volume degrees of
freedom of M5 branes sitting at the singular point \cite{orbifold}, 
which allows us to keep track of the orbifold physics 
without worrying about how to quantize M-theory.

For the moment, let us first concentrate on anomaly 
inflow mediated by $C$ field. Early studies taught us that
the orbifold fixed point itself has to carry 
$-1$ unit of M5 brane charge \cite{dasgupta,orbifold,flux}. 
After taking this into account,
the net M5 brane charge (in the covering space) is $2N-1$, so
the anomaly inflow induced by the M5 brane charge would be\footnote{
One may worry that we are smearing out the -1 charge of the fixed 
point just as we smear out $2N$ charge from M5 brane. However, we
find no reason to treat -1 charge differently. For instance, 
there is a known case, say, in F-theory context, where an orientifold 
7-plane is realized as a bound state of 7-branes, each of which are 
U-dual to D-branes \cite{sen}.}
\be
\frac{1}{2}\left(-(2N-1)I_8-(2N-1)^3\;\frac{p_2({\cal N})}{24}\right).
\ee
The overall factor 1/2 takes into account the $Z_2$ modding of $R^5$.
In the covering space, all that happens is that the field strength $G$
and the normal bundle $\cal N$ are restricted to respect the orbifolding 
action. This is more suggestively written as
\be
-\left(\left(N-\frac{1}{2}\right){\cal J}_8 
+N(2N-1)(2N-2)\,\frac{p_2({\cal N})}{24} \label{hmm'}
\right),
\ee
where we used the identity (\ref{sum}).

However, this cannot be the end of story. In particular, the correct
anomaly inflow must vanish when $N=0$, since there is no worldvolume
anomaly to cancel in that case. Also $N=1$ should  correspond to a free 
(2,0) tensor theory, whose anomaly is ${\cal J}_8$. The above inflow
is clearly incapable of cancelling this worldvolume anomaly. There must
be additional inflow. 

A well-known fact is that failure of diffeomorphism invariance may occur
at fixed points in a background that is otherwise smooth \cite{horava1}. 
In other words, there is an additional anomaly inflow that is
associated with the geometry of the orbifold fixed point. 
Such contribution would have little to do with the topological terms,
$C\wedge I_8$ or $\tilde C\wedge d\tilde C\wedge d\tilde C$, 
and will generate new kind of anomaly inflow associated with the
singular nature of the fixed point. One source of anomaly is from the fact
that the dimensional reduction of the supergravity is itself anomalous 
and contributes six-dimensional anomaly localized at the singularities.
In the following, we will try to deduce such additional inflow from 
consistency of an orbifold compactification.

For this, consider anomaly associated with the compactification of M-theory on 
$T^5/Z_2$ with 32 fixed points. It has been observed that
six-dimensional gravitational anomaly is cancelled 
satisfactorily if each fixed point carries $-1$ of M5 brane charge and  
if we include 16 pairs of M5 branes stuck at half of 32 fixed points 
\cite{orbifold,dasgupta}. Let us work backward from this knowledge, then, 
and ask what conclusion we can draw on the contribution associated with 
the singularity of the fixed point when we include the $SO(5)$ anomaly 
in the discussion.

Using extension of FHMM adopted above (\ref{hmm'}), the inflow mediated by
$C$-charge is
\be
\frac{1}{2}\left(I_8({\cal T}\oplus{\cal N})+\frac{p_2({\cal N})}{24}\right)
=\frac{{\cal J}_8}{2},
\ee
for each of 16 orbifold points without M5 brane ($N=0$). 
For a fixed point with an M5 brane pair ($N=1$), on the other 
hand, the world-volume tensor multiplet contributes additional 
${\cal J}_8$, while the anomaly inflow changes the sign, 
\be
{\cal J}_8+\frac{1}{2}\left(-I_8({\cal T}\oplus{\cal N})-
\frac{p_2({\cal N})}{24}\right)={{\cal J}_8}
-\frac{{\cal J}_8}{2}=\frac{{\cal J}_8}{2}.
\ee
Thus, for either kind of fixed point, we find combined anomaly of amount 
${\cal J}_8/2$  from $C$-induced inflow and from worldvolume contribution.
The net anomaly must be zero, which happens only if further contribution 
of $-{\cal J}_8/2$ exists at each orbifold fixed point.

\vskip 1cm
\hskip 0mm
\begin{tabular}{|l|c|c|}\hline
Anomaly contribution  & fixed point & fixed point with a M5-brane pair \\ 
\hline
Tensor multiplet on M5        & 0 & ${\cal J}_8$ \\ \hline
Inflow mediated by $C$ & ${\cal J}_8/2$   & $-{\cal J}_8/2$  \\ \hline
{Inflow due to singularity }  & $-{\cal J}_8/2$ &$-{\cal J}_8/2$ \\ \hline
\end{tabular}
\vskip 0.5cm
\begin{quote}
{\bf Table 1.} The first row is due to one-loop contribution from worldvolume
fields. The second row is the anomaly inflow from spacetime
action involving $C$-field.
Finally, we used anomaly cancellation on the  $T^5/Z_2$ orbifold to
determine  the last row. See next section for an independent estimate
of the last row.
\end{quote}

\vskip 0.5cm
\noindent
This clearly indicates that we must assign  $-{\cal J}_8/2$ to each of
32 fixed points, which are all locally of type $R^5/Z_2$.

Coming back to $N$ pairs of M5-branes on $R^5/Z_2$, this implies that
the additional anomaly at origin of $R^5/Z_2$ should be $-{\cal J}_8/2$. 
Net inflow is then
\be
-\left(\left(N-\frac12\right){\cal J}_8 
+N(2N-1)(2N-2)\,\frac{p_2({\cal N})}{24}
\right)-\frac{{\cal J}_8}{2},
\ee
negative of which should be the anomaly of rank $N$ D-type theory. We
conclude that the anomaly of rank $N$ D-type theory is given by
\be
N{\cal J}_8 
+N(2N-1)(2N-2)\,\frac{p_2({\cal N})}{24}.
\ee
This does reduce to the free tensor theory anomaly when $N=1$, as it should.

\section{Untwisted Sector One-loop Anomaly}

However, this is not a rigorous computation. In particular, we
have not computed the anomaly inflow due to the singularity; rather
we surmised its value, $-{\cal J}_8/2$, by relying on the consistency 
of the $T^5/Z_2$ compactification. We computed the twisted sector
one-loop anomaly and $C$-induced anomaly, and then inferred what
contribution at the singularity is necessary to cancel these.
With no particular reason to doubt the consistency of the orbifold 
in question, any failure of the above procedure would be traced 
to the fact that we adopted and generalized the FHMM prescription to 
compute $C$-induced anomaly inflow, both in  the M5 realization of
(2,0) theories and in the $T^5/Z_2$ compactification.

More specifically, there are two potential issues. One is that the 
FHMM proposal itself involves higher order terms in the M-theory
action, which has not been derived independently.\footnote{ 
A different mechanism of anomaly cancellation
for a single M5 brane has been suggested \cite{bonora}.}
The other, more worrisome issue is
 whether the orbifolding might induce other changes
in the FHMM reasoning that we somehow missed. 
To the  same extent this prescription may be in 
doubt, so would be the inferred fixed point contribution, $-{\cal J}_8/2$,
and vice versa. An independent estimate of the latter would go a long way
in checking the overall validity of our computation.

Purely gravitational part of $-{\cal J}_8/2$, namely $-J_8/2$, 
was computed  directly from the untwisted sector \cite{orbifold,dasgupta}.
It arises at one-loop from untwisted sector fields, which is chiral in
the $T^5/Z_2$ orbifold case.\footnote{ 
Since the initial 11-dimensional supergravity
is anomaly free, failure of diffeomorphism invariance must localize to 
orbifold fixed points. A similar
phenomenon occurs in the Horava-Witten realization of heterotic string,
i.e., M-theory compactified on $S^1/Z_2$ \cite{horava1,horava2}. 
The dimensional reduction of 11-dimensional supergravity multiplet produces 
10-dimensional  chiral supergravity multiplet. 
Anomaly of the latter is distributed equally at the two fixed points and 
is cancelled by the twisted sector anomaly from the Yang-Mills multiplets
in the two  $E_8$'s.}  We must at least generalize this computation to
include axial one-loop anomaly from untwisted sector.
One might hope that the required anomaly, $-{\cal J}_8/2$, at 
each fixed point is explained entirely by the one loop contribution.
We will see shortly that this is the case up to a (yet to be identified)
local counter term.

When one compactifies the 
11-dimensional supergravity on $T^5/Z_2$,  dimensional
reduction produces the following 
field contents:  one graviton and fifteen scalars from the metric; 
four anti-chiral gravitino and twenty chiral spinors from the
11-dimensional gravitino; ten scalars, five chiral 2-forms, and 
five anti-chiral 2-forms from the 3-tensor $C$. With respect to
the unbroken $(2,0)$ supersymmetry, they group into a supergravity
multiplet and five tensor multiplets.

To compute the one-loop anomaly, we also need to
know the $SO(5)$ representations under which these untwisted sector fields
fall into. With respect to the $SO(5)$ R-symmetry, which is nothing but the 
rotational group in the compactified direction, the representations are
\bea
g_{\mu\nu} &\rightarrow & 1,\nn
\psi_\mu^{s(-)} &\rightarrow& 4,\nn
B_{i\mu\nu}^{(-) }&\rightarrow &5;\\
&&\nn
B_{i\mu\nu}^{(+) }&\rightarrow &1\otimes 5,\nn
\psi^{s(+)}_i&\rightarrow & 4\otimes 5,\nn
\phi_{ij}&\rightarrow &5\otimes 5=(14\oplus 1)\oplus 10.
\eea
In other words, in addition to acting within each supermultiplet as R-symmetry
usually does, $SO(5)$ also rotates the five tensor multiplets.
This can be seen by observing that chiral and anti-chiral tensors,
$B_{i\mu\nu}^{(\pm)}$, combines to form 
$C_{i\mu\nu}$, or by observing that out of the 25 scalars,
$(14+1)$ come from the metric and 10 come from the Hodge dual of $C$ along 
$T^5/Z_2$.

To the one-loop anomaly, contributions from the chiral and anti-chiral
tensors cancel away, and the 
nontrivial contribution arises only from fermionic
fields. Following Alvarez-Gaume and Witten \cite{anomaly}, we find that 
the anomaly polynomial is the eight-form part of
\be
-\frac{1}{2}\,{\cal A}({\cal T})\wedge (ch_V({\cal T})-1)
\wedge ch_4({\cal N})+
\frac{1}{2}\,{\cal A}({\cal T})\wedge ch_{4\otimes 5}({\cal N}),
\ee 
where $ch_{\cal R}({\cal E})$ is the Chern-character of the bundle $\cal E$
evaluated on representation ${\cal R}$, and ${\cal A}$ is the 
${\cal A}$-genus associated with the Dirac operator on spinor. 
A direct evaluation gives,
\be
\frac{1}{3}\left( p_2({\cal T})-p_2({\cal N})-\frac{(p_1({\cal T})
-p_1({\cal N}))^2}{4}\right)-\frac{1}{2}\,p_1({\cal N})\wedge \left(
p_1({\cal T})-p_1({\cal N})\right), \label{untwisted}
\ee
which contributes
\be
-\frac{1}{2}{\cal J}_8-\frac{1}{64}\;p_1({\cal N})\wedge \left(
p_1({\cal T})-p_1({\cal N})\right),
\ee
to each of 32 fixed points. 

The first term reproduces 
{\it the anomaly inflow due to the singularity  } of  Table 1.,
\be
-\frac{1}{2}{\cal J}_8,
\ee
but we find a disagreement of amount,
\be
-\frac{1}{64}\;p_1({\cal N})\wedge \left(
p_1({\cal T})-p_1({\cal N})\right),
\ee 
from the second term.
The discrepancy may not  be as bad as it appears, however. It does reproduce 
both purely gravitational and purely axial part. In particular,
the part proportional to $p_2({\cal N})$ is consistent with the 
previous estimate;  we set out to check the validity of the
FHMM-like prescription producing the irreducible term, $p_2({\cal N})$, 
and found that this part of anomaly is reproduced satisfactorily by
untwisted sector one-loop computation.

The remaining piece is in a product form and could be cancelled away by a local
counter term  on the fixed point. A priori, what we computed above is
only one specific type of contribution from untwisted sector.
Since the discrepancy seems to lie entirely at the orbifold fixed point, we
suspect that the failure here has something to do with not taking
into full account of the orbifold geometry.
One property of the  fixed point we have not used properly
is that the space $R^5/Z_2$ has the nontrivial Stieffel-Whitney class
$w_4$, or equivalently $p_1/2$ considered as an element of the $Z_2$ 
cohomology. 
While this fact is related to the $-1$ charge of the fixed point,
nontrivial $w_4$ should have its own physical consequences. It is
conceivable that such a twist generates a counter-term, either via a 
boundary term in the spacetime action or via a mechanism similar to
FHMM but associated with curvature.

\section{Conclusion}

Adopting an extension of FHMM and HMM,  and employing the
consistency of $T^5/Z_2$ orbifold of M-theory, we argued that the
anomaly polynomial of rank $N$  D-type $(2,0)$ theory is
\be
N{\cal J}_8 
+N(2N-1)(2N-2)\,\frac{p_2({\cal N})}{24}.
\ee
An independent check was performed by computing 
one-loop untwisted sector contribution in M-theory compactified on $T^5/Z_2$.
Modulo a term cancellable by a local counter-term, it supports the above
anomaly estimate.

Recently, K. Intriligator put forward an interesting 
conjecture \cite{intriligator} that
the general anomaly polynomial of the ADE $(2,0)$ theories takes
the form,
\be
r{\cal J}_8+g\times c_2\times\frac{p_2({\cal N})}{24},
\ee
where $r$, $g$, $c_2$ are the rank, the dimension, the dual Coxeter number
of the respective Lie algebra. As noted by Intriligator, the FMM estimate 
of the anomaly for A theories indeed agrees with the conjecture.
The rank $r=N$ D-type algebra is $so(2N)$, whose 
dimension is $g=N(2N-1)$ and whose dual Coxeter number is $c_2=2N-2$, 
so the result for D-type theories here also  agrees with the
conjecture. Microscopic origin of order $N^3$ anomaly 
\cite{klebanov,henningson1,henningson2} has been of
considerable interest but still remains mysterious. We hope that 
one might gain insight from the particular algebraic structure of its 
coefficient, advocated by Intriligator and confirmed in part here.

An independent confirmation of the result might be available in the 
AdS/CFT setting.
A. Tseytlin \cite{tseytlin} attempted to compute sub-leading contributions 
to conformal anomaly, which is presumably connected to the gravitational
and axial anomaly via supersymmetry. His computation revealed no
conflict with Ref.~\cite{hmm}, but the agreement was somewhat
partial since only some of order $N$ terms are computed explicitly.
One may hope that a refined version of 
such computation will give a powerful, albeit indirect, check 
of axial anomaly computation of both A-type and D-type theories;
In the AdS/CFT picture, the dual spacetime
backgrounds are devoid of any singularity; the two cases
differ only by the transverse four-manifold being either $S^4$ or $RP^4$.
This suggests that no extra difficulty would arise for D-type theories.

Similar
computation of anomaly for E-type theories would be desirable, but
this case seems much more elusive. No brane picture  is known to exist,
while a useful AdS/CFT picture is also difficult to come in part
by due to the lack 
of large $N$ limit. A type IIB picture may work better for a universal 
approach to all ADE cases, but it remains an open problem.

\vskip 1cm
I am grateful to Sangmin Lee for an independent check of Eq.~(\ref{untwisted})
and to Gabriele Ferretti for pointing out a notational error in an earlier 
version. I would like to thank Theory Group at University of Chicago,
where bulk of the manuscript was written, for warm hospitality.

\end{document}